\documentclass[prl,amsmath,amssymb,twocolumn,showpacs,amsfonts,twocolumn]{revtex4}
\usepackage{graphicx}

\def \beq {\begin{equation}}
\def \eeq {\end{equation}}

\begin{document}
\title{Quantum Zeno Effect in Radical-Ion-Pair Recombination Reactions}
\author{I. K. Kominis}

\affiliation{Department of Physics, University of Crete, Heraklion
71103, Greece} \affiliation{Institute of Electronic Structure and
Laser, Foundation for Research and Technology, Heraklion 71110,
Greece}

\date{\today}

\begin{abstract}
Radical-ion pairs are ubiquitous in a wide range of biochemical reactions, ranging from photosynthesis to magnetic sensitive
chemical reactions underlying avian magnetic navigation. We here show that the charge recombination of a radical-ion-pair is a continuous
quantum measurement process that interrogates the spin state of the pair. This naturally leads to the appearance of the quantum
Zeno effect, explaining a large amount of data on unusually long-lived radical-ion-pairs.
\end{abstract}
 \maketitle Radical-ion pairs
are playing a fundamental role in a series of biologically
relevant chemical reactions, ranging from charge transfer
initiated reactions in photosynthetic reaction centers
\cite{balabin} to magnetic sensitive reactions abounding
in the field of spin-chemistry \cite{timmel}, and in particular in
the biochemical processes understood to underlie the biological
magnetic compass of several species having the ability to navigate
in earth's magnetic field \cite{schulten,ritz}.

\begin{figure}[!h]
\centering
\includegraphics[width=5 cm]{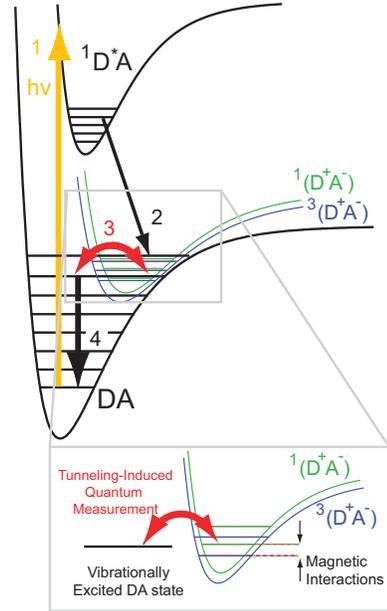}
\caption{Generic level structure and recombination dynamics in a radical-ion-pair, taking place in four steps: 1, photoexcitation,  2, RIP creation, 3, tunneling-induced quantum measurement of RIP's spin state and 4, final decay to the ground state.}
\label{fig:potentials}
\end{figure}

In Fig. \ref{fig:potentials} we depict a generic model for the
radical-ion-pair (RIP) creation and recombination dynamics. A
donor-acceptor molecule DA is photo-excited (D$^{*}$A) and a
subsequent charge-transfer creates the RIP (D$^{+}$A$^{-}$).
The singlet and triplet states of the RIP ($^{1}$D$^{+}$A$^{-}$,
$^{3}$D$^{+}$A$^{-}$) are split by internal magnetic interactions of the RIP's two unpaired electrons
with external magnetic fields and internal hyperfine couplings. The
RIP is initially created in the singlet state, which is not an
eigenstate of the magnetic Hamiltonian, and therefore a
singlet-triplet (S-T) coherent mixing commences. The RIP
eventually tunnels into an excited state of the neutral recombined
molecule DA, which quickly decays into the ground state.
As is well known \cite{jortner,kobori}, electron transfer in RIP recombination reactions is
fundamentally a quantum-mechanical tunneling process.
In this Letter we will show that this process constitutes a continuous quantum
measurement of the RIP's spin state. Like every quantum
measurement, this one is no exception to the rule that
measurements performed on a quantum system lead to decoherence \cite{braginsky}.
However, under appropriate conditions involving the measurement
rate and the intrinsic frequency scale of the RIP, the quantum
Zeno effect \cite{mishra} appears and leads to two physically
significant consequences: (i) dephasing rates of the RIP S-T
coherent mixing are suppressed, and (ii) the RIP's spin state is
delocalized, i.e. there is a high probability of triplet state
occupation even if the singlet and triplet states are
non-degenerate.

Quantum Zeno effects appear in several physical systems, some of
which are very similar to radical-ion-pairs, like the ortho-para
conversion in molecular spin isomers \cite{ortho}, ultra-cold atom tunneling through optical potentials \cite{raizen}, or the
suppression of transverse spin-relaxation due to spin-exchange
collisions in dense alkali-metal vapors \cite{happer_tang,kominis_pla}. In the
latter case, atomic spin-exchange collisions, of the form
$\mathbf{s}_{1}\cdot\mathbf{s}_{2}$, where $\mathbf{s}_{1}$ and
$\mathbf{s}_{2}$ are the electron spins of the two colliding
atoms, probe the atomic spin state. When the collision rate
(measurement rate) exceeds the intrinsic frequency scale of the
system, which is the Larmor frequency of spin precession in the
applied magnetic field, the effective decay rate of the spin
coherence is suppressed, a phenomenon that has led to the
development of new ultra-sensitive atomic magnetometers
\cite{romalis_prl}. The RIP tunneling into the neutral state is
essentially a scattering process \cite{carminati}, not
unlike atomic collisions, that performs a measurement of the RIP's spin
state, since tunneling can only proceed if the RIP is
in the singlet spin state. Quantum Zeno effects have been extensively
analyzed in the literature \cite{pascazio_review,pascazio,kaulakys,shimizu}, both with respect to pertaining physical systems,
as well the general conditions leading to the quantum Zeno effect or its inverse, the
anti-Zeno effect \cite{kurizki1,pascazio_prl,kurizki2}.

In the following, we are going to capitalize on the remarkably
strong analogy between radical-ion-pairs and yet another physical
system, namely two coupled quantum dots \cite{gurvitz,sw,wiseman,goan,oxtoby}
being continuously interrogated by a point contact. We are going
to identify the analogous physical observables of the two systems
and then setup the corresponding evolution equation describing the
RIP state, in order to directly arrive at the basic physical
results. The electron hopping between the two dots is the analog
of the S-T coherent mixing taking place in the RIP, whereas the
measurement performed by the point contact corresponds to the
spin-state-dependent RIP tunneling into an adjacent excited state
of the recombined DA molecule.

We will consider the simplest possible RIP model, in which the
triplet-state manifold is degenerate (and defines the zero
energy), and the singlet state has energy $\omega$:
\beq
{\cal H}_{\rm RIP}=\omega c_{S}^{\dagger}c_{S}+\Omega(c_{S}^{\dagger}c_{T}+c_{T}^{\dagger}c_{S})
\eeq
where $\Omega$ is the S-T mixing frequency. The tunneling to a
nearby excited DA state $|a\rangle$ with energy $\omega_{a}$ can only occur if the RIP is in the
singlet state, hence the tunneling Hamiltonian is
\beq
{\cal H}_{\rm T}=T_{\rm Sa}\, c_{\rm S}^{\dagger}
a^{\vphantom{\dagger}}+T_{\rm Sa}^*\, c_{\rm S}^{\vphantom{\dagger}}
a^{\dagger},  \label{th}
\eeq
where $T_{\rm Sa}$ is the tunneling amplitude. This Hamiltonian embodies angular momentum conservation in the tunneling process, i.e. the
tunneling amplitude for the triplet state is zero. In reality, the tunneling Hamiltonian is more complicated, since there are several
resonant vibrational states $|a\rangle$, and the recombination rate is given by \cite{jortner}
\beq
k=(2\pi/\hbar)|V|^{2}\sum_{a}{|f_{S,a}|^{2}\delta(\omega-\omega_{a})}
\eeq
where $V$ is the electronic matrix element and $f_{S,a}$ the vibrational overlap between the nuclear wavefunctions of $|a\rangle$ and the
singlet state of the RIP. In this realistic case, the tunneling Hamiltonian becomes
\beq
{\cal H}_{\rm T}=\sum_{a}{T_{\rm Sa}\, c_{\rm S}^{\dagger}
a^{\vphantom{\dagger}}+T_{\rm Sa}^*\, c_{\rm S}^{\vphantom{\dagger}}
a^{\dagger}},  \label{th}
\eeq
where now $T_{\rm Sa}=Vf_{\rm S,a}$.

Finally, the Hamiltonian of the DA excited state will be ${\cal H}_{\rm a}=\omega_{a}a^{\dagger}a$.
The operators $c_{S}$ ($c_{S}^{\dagger}$), $c_{T}$ ($c_{T}^{\dagger}$) and $a$ ($a^{\dagger}$) are electron annihilation (creation) operators
for the single-electron states $|S\rangle$, $|T\rangle$ and $|a\rangle$, respectively.
The rate constant $k$ is
termed the recombination rate, and will be later identified with
observable rate constants. The complete interaction Hamiltonian
governing the time evolution of the combined system is then ${\cal
H}={\cal H}_{\rm a}+{\cal H}_{\rm RIP}+{\cal H}_{\rm T}$. The same set of Hamiltonians has already been treated at \cite{sw}. In similar fashion,
it is readily shown that by tracing out the $|a\rangle$
degrees of freedom, we arrive at the dissipative evolution of the
RIP density matrix $\rho$: \beq {{d\rho}\over {dt}}=-i[{\cal
H}_{\rm RIP},\rho]-k{\cal D}[c_{S}^{\dagger}c_{S}]\rho \eeq where
the super-operator ${\cal D}[B]$ acts on the density matrix $\rho$
according to \beq {\cal D}[B]\rho=B^{\dagger}B\rho+\rho
B^{\dagger}B-2B\rho B^{\dagger} \eeq The occupation number
$c_{S}^{\dagger}c_{S}$ can also be written as
$Q_{S}=1/4-\mathbf{s}_{1}\cdot\mathbf{s}_{2}$, which is the
singlet-state projection operator in a RIP with the unpaired electron spins being
$\mathbf{s}_{1,2}$. In other words, the eigenvalues of $Q_{S}$ are 1 (RIP in the singlet state) and 0 (RIP in the triplet state). Since
$Q_{S}^{\dagger}=Q_{S}$ and $Q_{S}^{2}=Q_{S}$, we arrive at the
evolution equation \beq {{d\rho}\over {dt}}=-i[{\cal H}_{\rm
RIP},\rho]-k[Q_{S},[Q_{S},\rho]]\label{eq:ev} \eeq This is exactly
the evolution equation that follows from standard quantum
measurement theory \cite{braginsky,steck}, when the measured observable
is $Q_{S}$ and the measurement rate is $k$. We can generalize this by opening a
triplet recombination channel, with a recombination rate $k_T$ (the singlet recombination rate, so far denoted by $k$ is
renamed $k_{S}$). It is readily shown (since $Q_{S}+Q_{T}=1$) that the evolution equation is again given by (\ref{eq:ev}), with
$k=k_{S}+k_{T}$.

In the simple two-dimensional RIP model we are considering, the
density matrix, the Hamiltonian and the singlet-state projection
operator are $2\times 2$ matrices: \beq \rho=\begin{pmatrix}
\rho_{SS} & \rho_{ST}\\\rho_{TS} & \rho_{TT}\end{pmatrix},~{\cal
H}_{\rm RIP}=\begin{pmatrix} \omega & \Omega\\\Omega &
0\end{pmatrix},~Q_{S}=\begin{pmatrix} 1 & 0\\0 & 0\end{pmatrix}
\eeq
\begin{figure}
\centering
\includegraphics[width=7 cm]{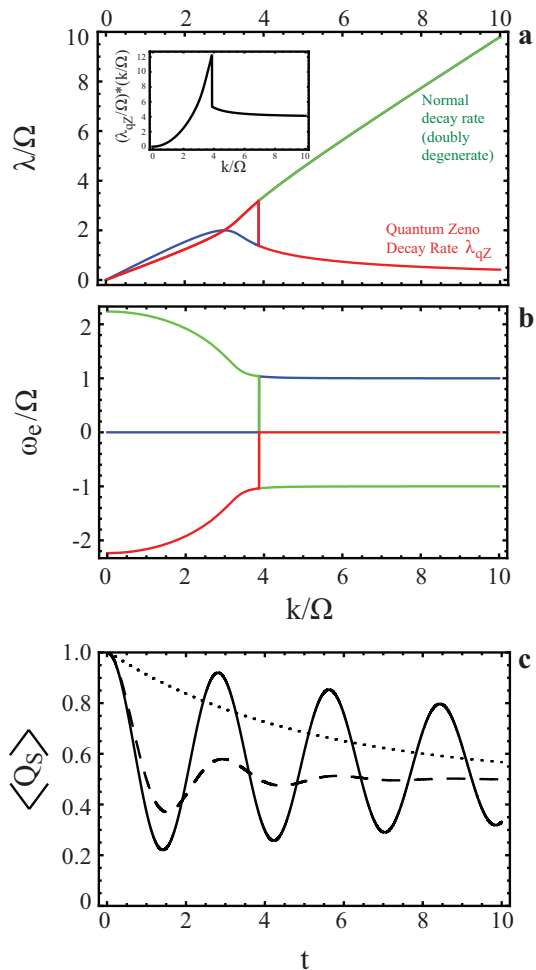}
\caption{Decay rates and eigenfrequencies of the evolution equation. (a) The "normal" decay rates are the ones that increase
with the measurement rate $k$. Quantum Zeno effect appears in the decay rate that is suppressed with $k$, denoted by $\lambda_{qZ}$, which is shown in the inset to be $\lambda_{qZ}\approx 4\Omega^{2}/k$ at high $k$.(b) Corresponding eigenfrequencies.The eigenfrequencies at high measurement rates are smaller than at low rates, an effect well-known
from spin-exchange relaxation collisions in dense alkali-metal vapors \cite{happer_tang}. In both (a) and (b) $\omega/\Omega=1$. (c) The expectation value $\langle Q_{S}\rangle$ for $\rho(t=0)=Q_{S}$, $\omega=\Omega=1$ and $k=0.1$ (solid line), $k=1$ (dashed line) and $k=20$ (dotted line).}
\label{fig:eig}
\end{figure}
The evolution equation (\ref{eq:ev}) can compactly be written as $d\rho/dt={\cal L}(\rho)$, where ${\cal L}$ is a super-operator.
The complex eigenvalues of the matrix $A$ resulting from the requirement ${\cal L}(\rho)=A\rho$ (where $\rho$ is now a column vector
$(\rho_{SS}~\rho_{ST}~\rho_{TS}~\rho_{TT})^{T})$ are of the form $-\lambda+i\omega_{e}$, where $\lambda\geq 0$ is
the decay rate and $\omega_{e}$ is the effective mixing frequency. In the simple system we are considering, there are three independent
density matrix elements, and hence three eigenvalues.
These are shown in Fig.\ref{fig:eig} as a function of the
recombination rate $k$ (normalized to the unperturbed mixing
frequency $\Omega$). Out of the three decay rates, two (degenerate ones) increase with $k$ and are termed "normal"; the quantum Zeno effect is manifested in two ways: (i) the third decay rate, $\lambda_{qZ}$, is suppressed with increasing measurement rate $k$, and (ii) the effective mixing
frequency drops in the limit of strong measurement (large $k$).
The scaling of $\lambda_{\rm qZ}$ with the measurement rate $k$ is
$\lambda_{\rm qZ}\sim\Omega^{2}/k$. This scaling is typical of quantum Zeno effects \cite{kurizki1,streed,pascazio_review}, since the Zeno time $\tau_{Z}$ is in this case given by $\tau_{Z}^{-1}=\sqrt{\langle S|{\cal H}_{\rm RIP}^{2}|S\rangle-\langle S|{\cal H}_{\rm RIP}|S\rangle^{2}}=\Omega$,
i.e. $\lambda_{qZ}=\tau/\tau_{Z}^{2}$, where $\tau=1/k$. The long-lived
eigenvalue $\lambda_{\rm qZ}$ will determine the long-time behavior of the density
matrix elements. A typical value of the recombination rate is
$k\approx 1~{\rm ns}^{-1}$. For an unperturbed mixing frequency
$\Omega\approx 30~{\rm ns}^{-1}$ (this corresponds to a hyperfine
coupling of 10 G), we find that $\lambda_{\rm qZ}\approx 1~{\rm
\mu s}^{-1}$. The lifetime of the S-T mixing process
can thus be prolonged by several orders of magnitude.

This naturally explains several experimental
observations regarding long-lived radical-ion pairs
\cite{verhoven_review,verhoven1,fukuzumi}. It should be noted that
by "recombination" rate we refer to the fast-decaying eigenmodes
of the density matrix, i.e. those for which the decay rate scales
proportionally with $k$.
Time scales on the order of 10 ps-1 ns,
which govern the creation of the RIP through photoexcitation and
decay of the excited DA-state could so far not be fathomed with
long-lived radical-ion-pairs, i.e. since the D$^{*}$A-RIP and
RIP-DA energy differences are comparable, why doesn't the RIP
disappear at sub-ns time scales? This cannot be explained by the
presence of the "metastable" triplet state, since the mixing rates
with the singlet state are typically in the 10 ns timescale.
Therefore, even if the RIP is created in the triplet state, it
should disappear fast through T-S mixing and singlet-channel
charge recombination. The quantum Zeno effect naturally leads to
RIP lifetimes that extend even to the ${\rm \mu}$s timescale. The theoretical models that were used until now
\cite{steiner,hore1,schulten2,hore2} to describe RIP recombination
dynamics masked the presence of the quantum Zeno effect, since
they treated the tunneling process of the RIP and the subsequent
decay to the DA ground state (steps 3 and 4 in Fig.
\ref{fig:potentials}) with a single, phenomenological density
matrix equation $d\rho/dt=-i[{\cal H}_{\rm RIP},\rho]-k(\rho
Q_{S}+Q_{S}\rho)$. This equation accounts for the depopulation of
the singlet RIP state at a rate $2k$, i.e. the probability $S={\rm
Tr}\{\rho Q_{S}\}$ to find the RIP in the singlet state decays
exponentially at a rate $2k$, unavoidably inducing a simultaneous dephasing of
the S-T coherent mixing at the the high rate $k$. It is noted for completeness that
the aforementioned semi-classical density matrix equation leads to similar results
just in this particular two-dimensional toy-model of the RIP. As soon as we move to
a realistic description of the RIP, which involves at least an 8-dimensional density matrix (4 is the two-electron spin multiplicity and
2 the spin multiplicity of one nucleus with spin-1/2, which is the bare minimum needed to form a RIP supporting singlet-triplet mixing), all decay rates scale proportionally to $k$ and there is
no manifestation of the Zeno effect. This is not the case with the density matrix equation (\ref{eq:ev}).

The second physical consequence of the quantum Zeno effect in the
RIP recombination merits some discussion. The probability for
significant triplet state population is high, on the order of
unity, even for S-T energy differences $\omega>\Omega$. This
effect has been discussed in \cite{gurvitz}, and in the simple RIP
model we are considering, it is seen in the fact that $\rho_{SS}$
tends to 1/2 in the long-time limit, irrespective of the problem's
frequencies. This has tangible consequences, since in realistic
systems, the triplet state RIP can recombine to other chemical
products, or in cases of RIPs in solution, the D and A
molecules will eventually diffuse away. In both cases, a large
probability of populating the RIP triplet state will be evident in
the reduced yield of recombined DA molecules.

We will finally elaborate on step 4 of Fig.\ref{fig:potentials},
namely the decay to the ground state of the neutral DA molecule.
The continuous quantum measurement performed by the tunneling
process into the excited DA state will at times be interrupted
when there is a definite measurement outcome. This is described by
the quantum-jump approach \cite{plenio,sw} of the
quantum-trajectories description of dissipative quantum systems. When the outcome of measuring $Q_{S}$  is 1, that is, the RIP is in the singlet
state for sure, tunneling and decay to the DA ground state can proceed. This is
formally described by the conditional evolution of the RIP's
quantum state $|\Psi\rangle$,
\begin{align}\nonumber
|\Psi(t+dt)\rangle=|\Psi(t)\rangle&-idt{\cal H}_{RIP}|\Psi(t)\rangle\nonumber\\&-kdt(Q_{S}-\langle Q_{S}\rangle)|\Psi(t)\rangle\nonumber\\
&+dN\left(Q_{S}/\sqrt{\langle Q_{S}\rangle}-1\right)|\Psi(t)\rangle\label{qj}
\end{align}
The stochastic point process that takes the values 0 and 1, i.e. $(dN)^{2}=dN$, and $M[dN]=2k\langle Q_{S}\rangle dt$, where $M[.]$ represents the
mean over all possible realizations of the process. The first three term in (\ref{qj}) represent the no-jump deterministic evolution of the RIP state, while the last term describes the quantum jump that eventually occurs opening the possibility for charge recombination, after which the DA molecule can relax to its ground state. This way we have closed the excitation-recombination cycle pictured in Fig. \ref{fig:potentials}. Analogously to the tunneling current \cite{sw,goan} we can define a "charge-recombination current" $R_{c}=dN/dt=k\langle Q_{S}\rangle$. The recombination reaction rates can then be defined as the characteristic rates appearing in the two-time correlation function $G(\tau)=E[R_{c}(t)R_{c}(t+\tau)]$, which is similarly to \cite{sw,goan} shown to be given by $G(\tau)=k^{2}({\rm Tr}\{e^{{\cal L}\tau}\rho_{s,\infty}\}-1/4)$, where $\rho_{s,\infty}=Q_{S}\rho_{\infty}Q_{S}$ is the singlet projection of
the steady state density matrix (which in this case is $\rho_{\infty}=1/2$). Thus the reaction rates are the eigenvalues of ${\cal L}$ that we have calculated and plotted in Fig. 1.

In summary, starting from first principles, we have here demonstrated the
fundamentally quantum-mechanical nature of the radical-ion-pair recombination
process. This provides a natural explanation of several unusual
experimental findings in RIP chemical reactions. More important,
it is not inconceivable that quantum Zeno effects in
radical-ion-pairs could be found to be intimately involved with
the quantum-mechanical foundations of photosynthetic reactions.
\begin{acknowledgements}
I acknowledge helpful discussions with Dr. D. Anglos as well as helpful comments by the anonymous Referees.
\end{acknowledgements}

\end{document}